\documentclass[twocolumn,showpacs,amsmath,amssymb,showkeys,floatfix,prl]{revtex4}
\usepackage{amsfonts,amsmath}
\usepackage{graphicx}
\usepackage{dcolumn}
\usepackage{bm}

\begin{document}

\title{Atmospheric, long baseline, and reactor neutrino data constraints on $\theta_{13}$}

\author{J. E. Roa$^1$, D. C. Latimer$^2$,   and D. J. Ernst$^1$}

\affiliation{$^1$Department of Physics and Astronomy, Vanderbilt University,
Nashville, Tennessee 37235}

\affiliation{$^2$Department of Physics and Astronomy, University of Kentucky, Lexington, Kentucky 40506}

\date{\today}

\begin{abstract}
An atmospheric neutrino oscillation tool that uses full three-neutrino oscillation probabilities
and a full three-neutrino treatment of the MSW effect, together with an analysis of the K2K, MINOS, and CHOOZ data, is used to examine the bounds on
$\theta_{13}$. The recent, more finely binned, Super-K atmospheric data is employed. For $L/E_\nu\gtrsim 10^4$ km/GeV,
we previously found significant linear in $\theta_{13}$ terms. This
analysis finds $\theta_{13}$ bounded from above by the atmospheric data while bounded from below by CHOOZ. The origin of this result arises from data in the previously mentioned very long baseline region; here, matter effects conspire
with terms linear in $\theta_{13}$ to produce asymmetric bounds on $\theta_{13}$. Assuming CP conservation,
we find $\theta_{13}=-0.07^{+0.18}_{-0.11}$ (90\% C.L.).

\end{abstract}



\maketitle

The phenomenon of neutrino oscillations \cite{exp,k2k,atmo,atmor} has been observed in a variety of experiments: solar, long baseline (LBL)  reactor, 
atmospheric, and LBL accelerator experiments.  Including the constraint 
imposed by the CHOOZ reactor experiment \cite{choo}, one may quantitatively determine the 
three mixing angles and two mass-squared differences that parameterize three-neutrino phenomenology  \cite{cgg07}. An outstanding question is the 
value of the mixing angle $\theta_{13}$. Present analyses \cite{cgg07} yield $\vert\theta_{13}\vert \leq 0.15$. We examine 
the impact of small effects, particularly those linear in $\theta_{13}$ \cite{dlde05,dlde05b,ps04,choubey}, on extracting this small parameter from the 
data. We find that including the full three-neutrino oscillation probabilities and a full three-neutrino MSW calculation are important for determining this 
mixing angle.

Knowledge of $\theta_{13}$ is a particularly important part of neutrino oscillation phenomenology because its value sets the magnitude of possible CP 
violating effects and the size of effects that might be used to determine the neutrino mass hierarchy. There are presently three new reactor experiments 
under development which are designed to measure $\theta_{13}$, Daya Bay \cite{DYB}, Double CHOOZ \cite{dCHO}, and RENO \cite{RENO}, as well as two long 
baseline experiments, T2K \cite{T2K} and NOvA \cite{nova}. The subsequent generation of experiments, e.g., those which will ascertain the level of CP 
violation, cannot proceed until the current generation better determines the value of $\theta_{13}$.

The standard model of neutrinos conserves flavor, as is required by the data. Neutrino oscillations require adding
{\it a posteriori} a mass term to the standard model Lagrangian. The standard model Lagrangian is diagonal in flavor; the
added mass term is diagonal in the basis which governs vacuum propagation. The relation between 
the two bases is given by a phenomenological unitary matrix $U_{\alpha i}$. In the absence of CP violation, it is real. We employ the standard 
representation \cite{pmns} written in terms of the three mixing angles, $\theta_{12}$, $\theta_{23}$, and $\theta_{13}$. In vacuum,
the probability that a neutrino of flavor $\alpha$ and energy $E_\nu$ will be detected a distance
$L$ from the source as a neutrino of flavor $\beta$ is given by
\begin{equation}
{\mathcal P}_{\alpha \beta}(L/E_\nu) = \delta_{\alpha \beta} - 4 
\sum^3_{\genfrac{}{}{0pt}{}{k <j,}{j,k=1}} (U_{\alpha j} U_{\alpha k} U_{\beta k} 
U_{\beta j}) \sin^2 {\varphi_{jk}}
\label{exact}
\end{equation}
with $\varphi_{jk} := 1.27\,\Delta_{jk}\,L/E_\nu$ and $\Delta_{jk}=m_j^2-m_k^2$, where $L$ is measured in km, $E_\nu$ in GeV, and the mass 
eigenvalues $m_i$ in eV.  If (anti-)neutrinos travel an appreciable distance through matter of sufficient density, then one must add to the 
Hamiltonian an effective potential to account for the (anti-)neutrino--matter interactions \cite{msw}.  This potential yields different effective mixing angles and neutrino masses.  For neutrinos which 
propagate through the earth over long baselines, it is crucial to take such matter effects into account.
We do so by using the approach developed in Ref.~\cite{ohsn}.  We employ a two density model of the earth: a mantle of density 4.5 gm/cm$^3$ 
and a core of density 11.5 gm/cm$^3$ with  radius 3486 km. This approach incorporate the MSW effect into a three-neutrino framework 
without the use of approximate oscillation formulae.  We note that it is possible for parametric resonances
to occur when neutrinos pass between regions of differing densities \cite{parametric}; our treatment of matter effects automatically insures that 
such resonances are fully incorporated.

Atmospheric neutrino experiments are unique in that the baselines span several orders of magnitude making them sensitive to an 
enormous region of relevant parameter space.  However, the SK-atmospheric experiment is the most difficult to model as one must use the detected charged 
lepton to infer the direction and energy of the incident neutrino.  A complete description of our analysis tool can be found in Ref.~\cite{jesus}. 
Statistical errors are included in the chi-square function whereas systematic errors are accounted for by using the pull method \cite{pulls}. We include 
43 pulls, the most important being the overall flux normalization. We also include \cite{jesus} a simple model of the multi-ring events. For 
CHOOZ, K2K, and MINOS, we utilize standard analysis techniques \cite{jesus}.

We introduce the commonly used ``sub-dominant approximation'' to provide a comparison for the full three-neutrino treatment.
It arises from an expansion in the ratio of the mass-squared differences, $\Delta_{12}/\Delta_{32}$. The oscillation probabilities are then given by
\begin{eqnarray}
{\mathcal P}_{ee}&=&1-\sin^22\theta_{13}\,\sin^2\left(\varphi_{32}\right) \nonumber \\
{\mathcal P}_{e\mu}&=& \sin^2\theta_{23}\,\sin^22\theta_{13}\,\sin^2\left(\varphi_{32}
\right) \nonumber \\
{\mathcal P}_{\mu\mu}&=&1-4\cos^2\theta_{13}\,\sin^2\theta_{23}\,(1-\cos^2\theta_{13}\,\sin^2\theta_{23})
\nonumber \\
&&\times \sin^2\left(\varphi_{32}\right)\,\,.
\label{subd}
\end{eqnarray}
Additional correction terms can be added \cite{cgg07,expans}. We effect this approximation by setting $\Delta_{21}=0$ in our full three-neutrino code; 
in this treatment, matter effects will differ slightly from the approximations used by others.
\begin{figure}
\includegraphics[width=3in]{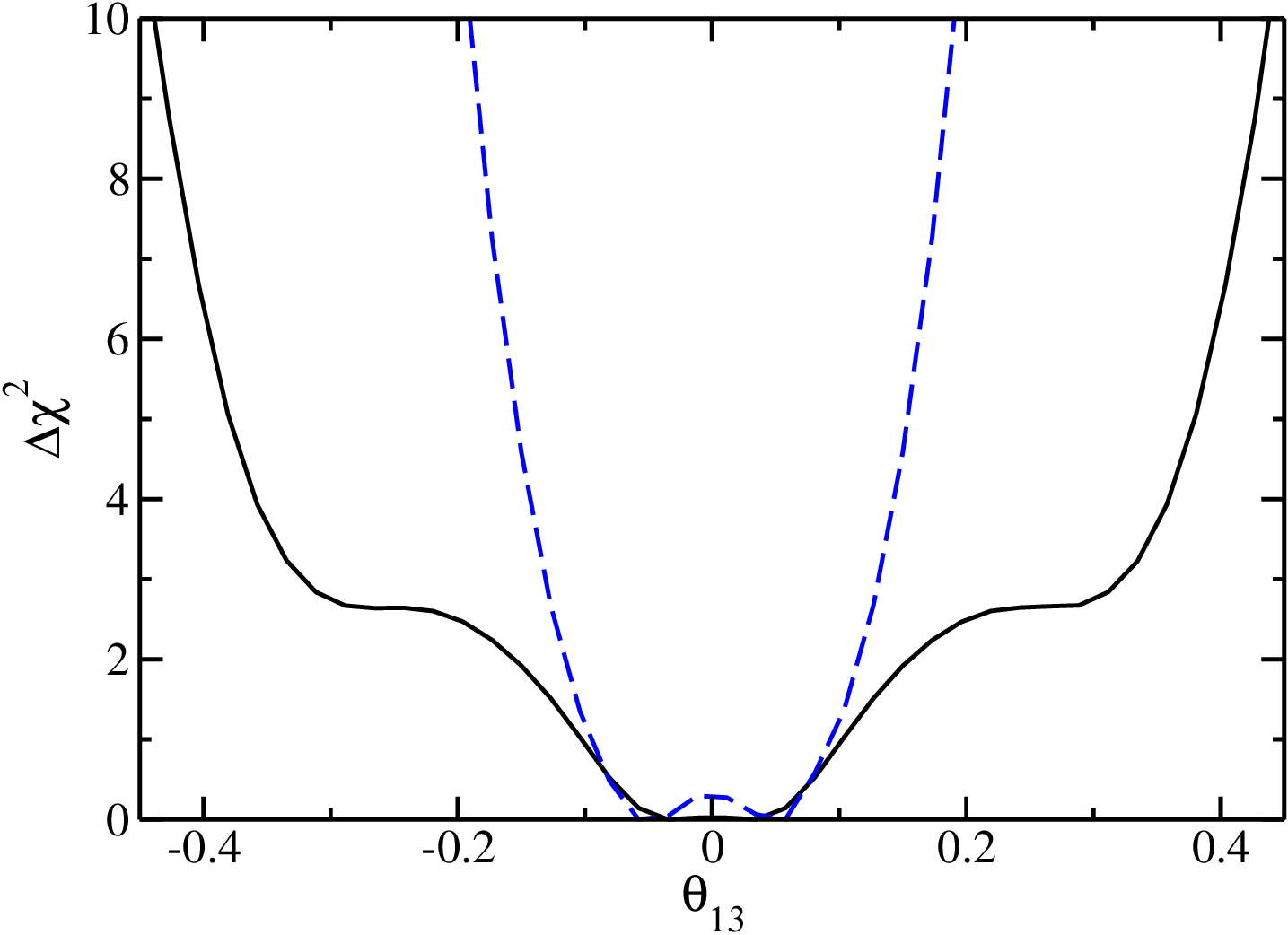}
\caption{[color online] $\Delta\chi^2$ versus $\theta_{13}$ in the sub-dominant approximation. The solid [black] curve utilizes only atmospheric data;
the dashed [blue] curve contains the atmospheric, LBL, and CHOOZ experiments.}
\label{fig1}
\end{figure}

We take the bounds on the mixing angles as $\theta_{13} \in [-\pi/2,\pi/2]$ and $\theta_{12},\theta_{23} \in [0,\pi/2]$, as
suggested in Ref.~\cite{dlde05c}. For no CP violation, this produces an allowed parameter space that is a single connected region.  The 
(equivalent) often used bounds, $\theta_{jk} \in [0,\pi/2]$ with Dirac CP phase $\delta=0, \pi$, produce two disconnected regions.

We begin utilizing the sub-dominant approximation. Ref.~\cite{atmor} showed that atmospheric data alone restrict the allowed value of 
$\theta_{13}$, although less so than does CHOOZ.
We use the data from Ref.~\cite{atmor}, which are binned more finely in energy than the original data \cite{atmo}.  In Fig.~\ref{fig1}, we plot 
$\Delta \chi^2$ versus $\theta_{13}$, varying $\theta_{23}$ and $\Delta_{23}$. The solid [black] curve contains only atmospheric data. Our results 
quantitatively reproduce those of Ref.~\cite{atmor} which also used the sub-dominant approximation.
Both give $\sin^2\theta_{13} < 0.14$ (or $\vert\theta_{13}\vert < 0.38$) for $\Delta\chi^2 < 4.6$.
Reproducing this result is a strong test for our analysis tool.  The dashed [blue]  curve in Fig.~\ref{fig1} represents $\Delta\chi^2$ for the data set 
which includes the atmospheric \cite{atmor}, LBL (K2K and MINOS  \cite{k2k}),
and CHOOZ \cite{choo} experiments.  It is CHOOZ which restricts $\theta_{13}$ much more so than the atmospheric data in 
the sub-dominant approximation.

\begin{figure}
\includegraphics*[width=3in]{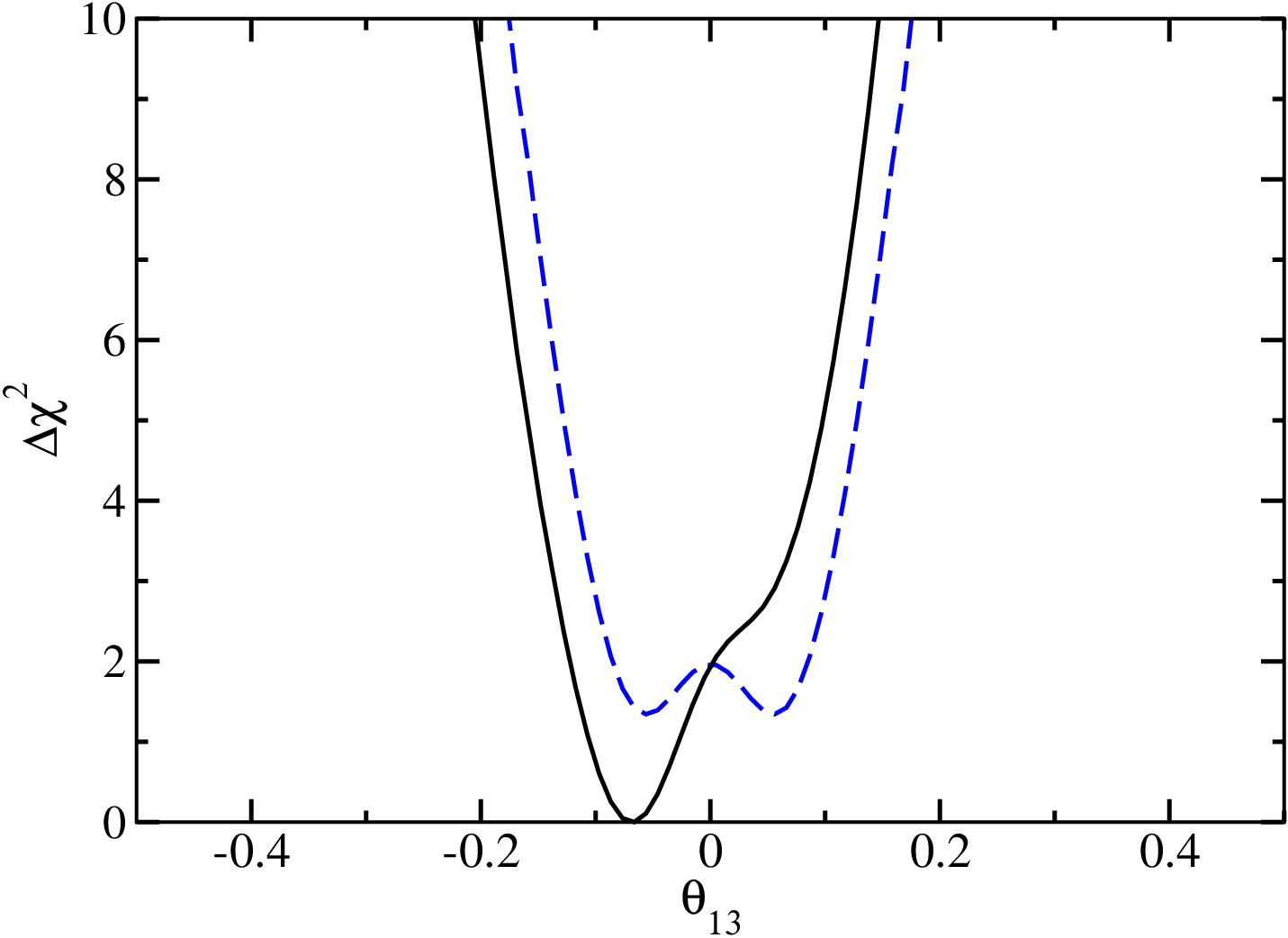}
\caption{[color online] $\Delta\chi^2$ versus $\theta_{13}$. The dashed [blue] curve includes all data and uses the sub-dominant approximation.  The solid [black] curve incorporates the same data with a full 
three-neutrino calculation. $\Delta\chi^2$ for both cuves is normed by the minimum value of the full three-neutrino calculation.}
\label{fig2}
\end{figure}

We now compare the sub-dominant approximation with the full three-neutrino calculation. In Fig.~\ref{fig2}, we plot $\Delta \chi^2$ for the full data set 
(atmospheric, LBL, and CHOOZ) using the sub-dominant approximation, dashed [blue] curve, and the full three-neutrino calculation, solid [black] curve.  We fix 
the solar parameters at their best fit values \cite{cgg07}, $\theta_{12}=0.58$ and $\Delta_{21}=8.0\times 10^{-5}$ eV$^2$, as 
this analysis largely decouples from these two parameters. Note that the sub-dominant results are symmetric about $\theta_{13}=0$, as is manifest in Eqs.~(\ref{subd}).  For the full three-neutrino model, the asymmetry about zero is primarily due to terms in the oscillation probabilities which 
are linear in $\theta_{13}$. We have previously investigated the importance of such terms for very long baselines,  $L/E \gtrsim 10^4$ km/GeV 
\cite{dlde05,dlde05b}. Such linear terms and the interference between the $\Delta_{21}$ and $\Delta_{32}$ oscillations 
have also been investigated in  Refs.~\cite{ps04,choubey}. The importance of this very long baseline region in  the atmospheric data has also been noted 
in Ref.~\cite{ghosh}.  Terms linear in $\theta_{13}$ are not negligible when extracting the value of this small angle.

\begin{figure}
\includegraphics*[width=3in]{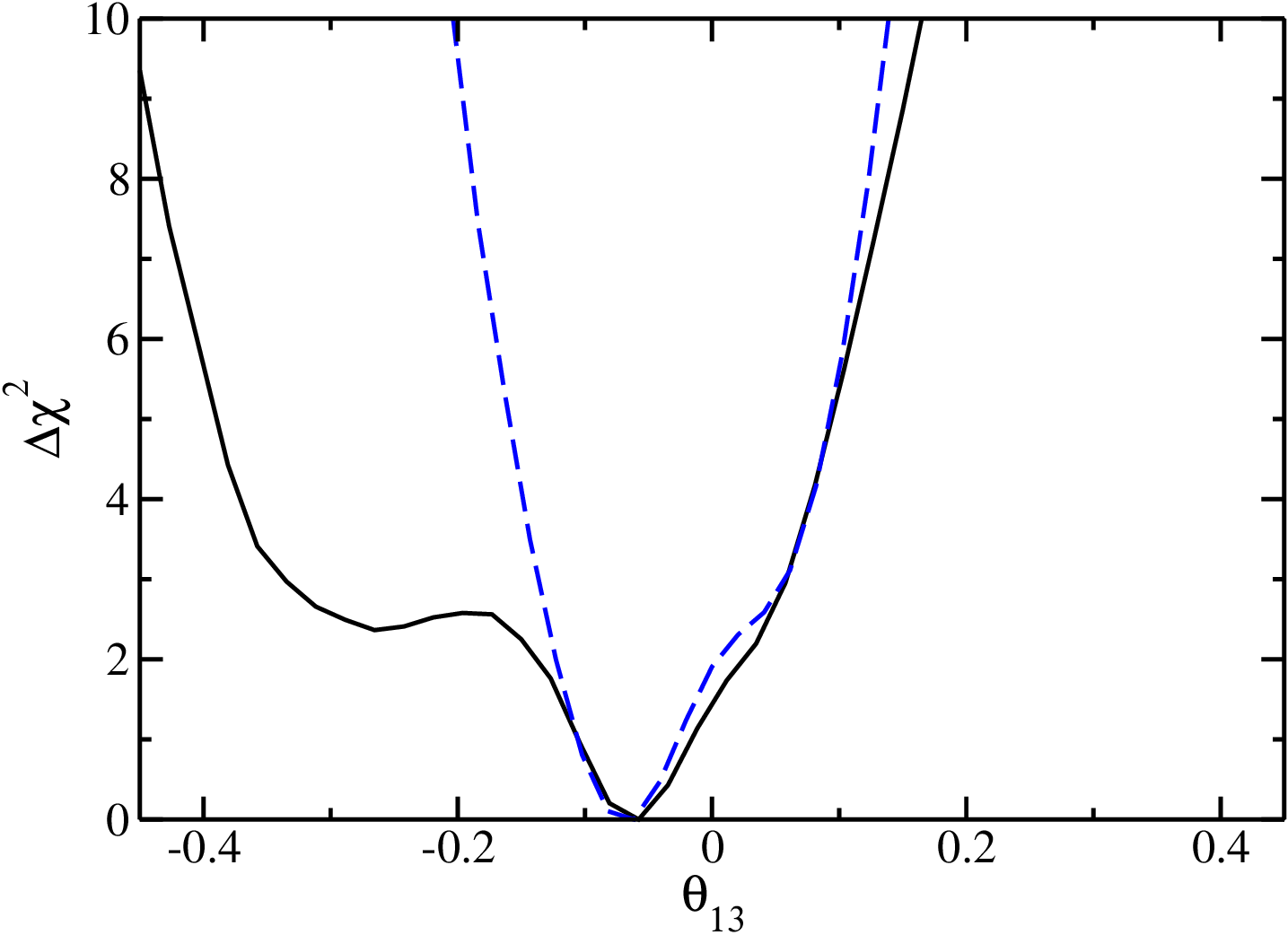}
\caption{[color online] $\Delta\chi^2$ versus $\theta_{13}$ using a full three-neutrino model. The solid [black] curve utilizes only atmospheric data; the dashed [blue] curve contains the atmospheric, LBL, and CHOOZ experiments.}
\label{fig3}
\end{figure}

We next examine the contribution of the atmospheric data alone  to $\Delta\chi^2$, the [black] solid curve in Fig.~\ref{fig3}.  The dashed [blue] curve employs the full data set. For positive $\theta_{13}$, the atmospheric data are more restrictive 
than even CHOOZ. The restrictions for negative $\theta_{13}$ are set by the CHOOZ data.
Overall, we find the allowed region for $\theta_{13}$
to be asymmetric about zero, bounded from above by atmospheric data and bounded from below by CHOOZ. The final value is $\theta_{13}=-0.07^{+0.18}_{-0.11}$ at 90\% confidence level, corresponding to $\Delta\chi^2=6.25$ for a 
three parameter analysis of this data set. In Fig.~\ref{fig3}, we,  as have others \cite{nzero}, find a non-zero value for $\theta_{13}$; furthermore, we find
a statistically insignificant preference for a negative value. 

Which subset of atmospheric data results in the strict upper bound on $\theta_{13}$ and the lack thereof from below? To answer this, we examine $\theta_{13}=\pm 0.15$ which has
$\Delta\chi^2\sim9$. We find that the sub-GeV fully contained events are responsible for two-thirds of this $\Delta \chi^2$. Furthermore, one-half of the total change in chi-squared (4.5) comes from the single angular bin, 
$-0.8 < \cos\vartheta < -0.6$, bin II, for fully contained $e$-like events, and the two lowest energy bins in which the charged lepton momentum, $p_\ell$, is  less than 400 MeV/c.
This is well into the very long baseline region mentioned previously where we expect contributions from terms  linear in $\theta_{13}$. Bin I, $-1.0 < \vartheta < -0.8$, contains neutrinos which traverse the core suppressing 
their amplitude of oscillation.

\begin{figure}
\includegraphics*[width=3in]{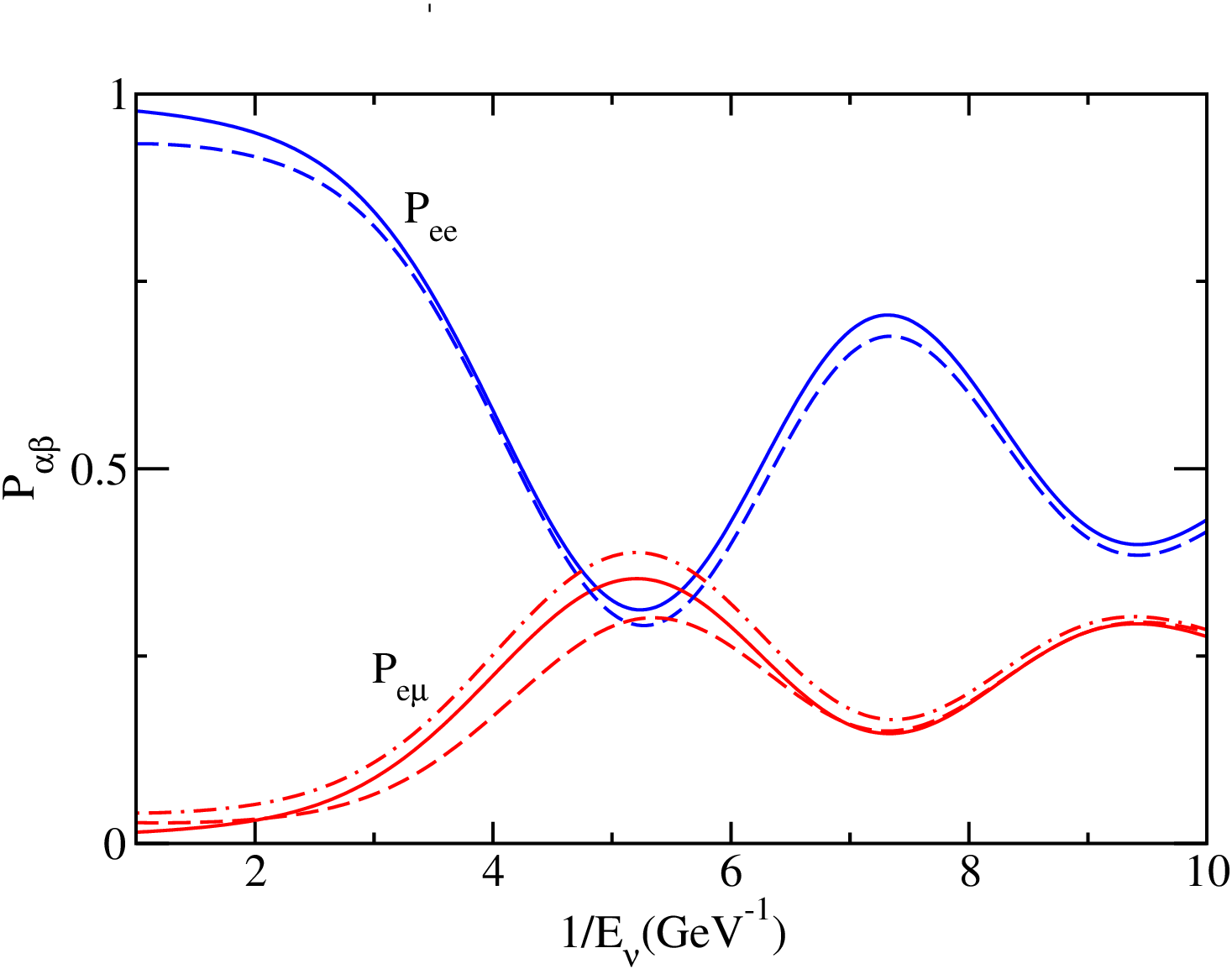}
\caption{[color online] The oscillation probabilities ${\mathcal P}_{ee}$ and ${\mathcal P}_{e\mu}$ versus the inverse neutrino energy $E_\nu^{-1}$. The probabilities have
been averaged over the angular bin II and folded with a 6\% error in the energy. The solid curves use the best fit values of
the parameters. The dash (dot-dash) curves are $\theta_{13}=+0.15\,\,(-0.15)$.}
\label{fig4}
\end{figure}

\begin{figure}
\includegraphics*[width=3in]{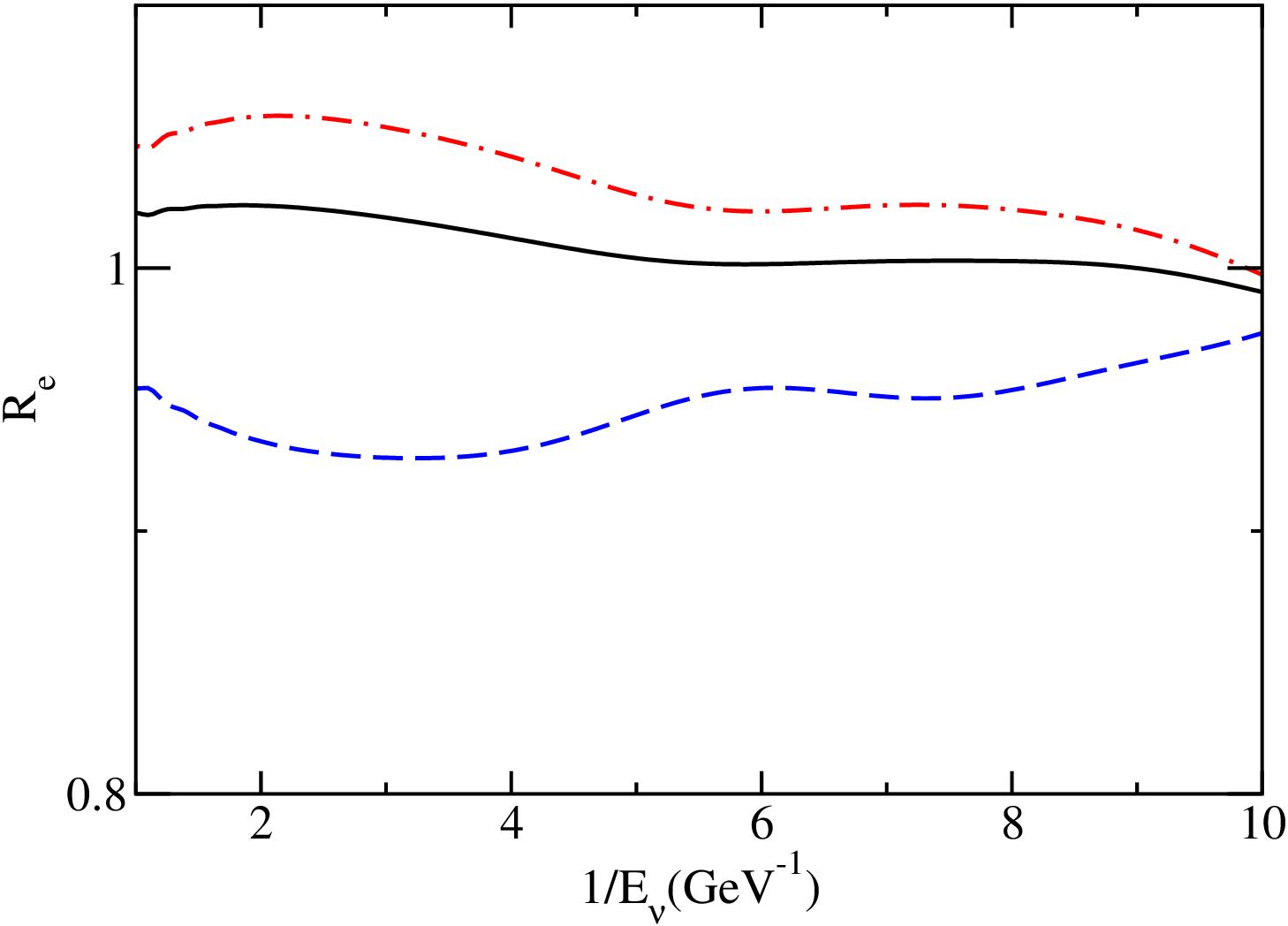}
\caption{[color online] The measured quantity $R_e$ versus the inverse neutrino energy $E_\nu^{-1}$ for angular bin II. 
The solid [black] curve utilizes the best fit parameters, the dashed [blue] curve $\theta_{13}=+0.15$, the dot-dashed [red] curve $\theta_{13}=-0.15$}
\label{fig5}
\end{figure}

In Fig.~\ref{fig4}, we plot oscillation probabilities 
${\mathcal P}_{\alpha\beta} (E_\nu^{-1})$ for angular bin II and the lowest energy bin, $p_{\ell}< 240$ MeV/c.
The solid curves use the best fit parameters, the dash (dot-dash) curves use $\theta_{13}=+0.15\,\,(-0.15)$. For ${\mathcal P}_{ee}$, the top two [blue] curves, there 
is only a quadratic term in $\theta_{13}$ which lowers ${\mathcal P}_{ee}$.  Examining the lower three [red] curves, we see that ${\mathcal P}_{e\mu}$ decreases (increases) with positive (negative)  $\theta_{13}$.   The number of 
e-like events measured in an atmospheric experiment is related to 
$R_e={\mathcal P}_{ee}+r{\mathcal P}_{e\mu}$ with $r$ the ratio of the $\nu_\mu$ to $\nu_e$ flux at the source. 
From Fig.~\ref{fig4}, effects due to terms linear in $\theta_{13}$ combine constructively for positive $\theta_{13}$ and destructively for negative $\theta_{13}$.
This is confirmed in Fig.~\ref{fig5} where we compare the effect of positive and negative $\theta_{13}$ upon $R_e$ for neutrinos in bin II.  Here, we also discover that the excess of e-like events at low energies \cite{ps04} 
results in the strict bound on positive values of $\theta_{13}$ in contrast to negative values.  The effect is enhanced by an MSW resonance near 
$E_\nu=100$ MeV for a mantle density of 4.5 gm/cm$^3$. Notice that the effect is for bins where the neutrino travels through almost the entire Earth, a distance such that the resonance is fully developed.
Previously, the constancy of $R_e$ was found to impose an upper bound on $\vert \theta_{13} \vert$ \cite{dharam}; here we see that $R_e$ also helps constrain the sign of 
this mixing angle.

Probing further, we may approximate the e-like events at Super-K for the very LBL sub-GeV data as in Ref.~\cite{dlde05b}
\begin{eqnarray}
R_e &\simeq& \nonumber\\
1&+&r \sin^2 2 \theta_{12}^m \left[\tfrac{1}{2}-\tfrac{1}{r}+ \cot({2 \theta_{12}^m}) \theta_{13} - \varepsilon
 \right] \sin^2 \varphi_{21}^m,~~~~ \label{re}
\end{eqnarray}
where the superscript $m$ refers to the effective parameter values in matter and $\varepsilon$ represents the deviation from maximal mixing, 
$\theta_{23} = \pi/4 + \varepsilon$.
Note the term linear in $\theta_{13}$ is proportional to $\cot 2\theta_{12}^m$. At the MSW resonance, $\theta_{12}^m =\pi/4$, and this mixing angle 
increases with energy up to $\pi/2$.  Thus the coefficient of the $\theta_{13}$ term in Eq.~(\ref{re}) is negative above the resonant energy so that a
negative $\theta_{13}$ produces an excess of e-like events.
The bounding of $\theta_{13}$ from above by the atmospheric data depends critically on incorporating the full MSW effect. For these low energy neutrinos, one has $r \simeq 2$; the dependence of our 
results upon this ratio is not severe. More significant is the dependence upon $\varepsilon$; the octant of $\theta_{23}$ can conspire to enhance or 
suppress the effect of $\theta_{13}$ upon an excess of e-like events.

In this new era of precision neutrino experiments, small effects, such as those arising from $\theta_{13}$, require a careful treatment. Future reactor experiments \cite{DYB,dCHO,RENO} are sensitive to $\theta_{13}^2$ and thus 
can determine the magnitude of $\theta_{13}$, but not its sign. The long baseline experiments, \cite{T2K,nova} will contain small effects that are linear in $\theta_{13}$, while an upgraded Super-K will produce additional data 
in the region most sensitive to effects linear in $\theta_{13}$. How these different data interplay with each other in determining this mixing angle will be most interesting. 

We find that present atmospheric data restrict the value of $\theta_{13}$ from above, while the limit from below remains as determined by CHOOZ, and that 
$\theta_{13}=-0.07^{+0.18}_{-0.11}$, assuming no CP violation. It is important to realize first that $\theta_{13}$ can be negative \cite{dlde05c} and, 
second, that linear effects lead to asymmetric errors. Our analysis requires the use of the more finely binned atmospheric data, Ref.~\cite{atmor}, the use 
of the full three-neutrino oscillation probabilities, and inclusion of the full MSW effect.

\section{ACKNOWLEDGMENTS}

The work of J.\ E.\ R.\ and D.~J.~E.~is supported, in part, by US Department of Energy Grant DE-FG02-96ER40975; the work of J.~E. R. is supported, in part by CONACyT, Mexico; the work of D.~C.~L.~is supported, in part, 
by US Department of Energy Grant DE-FG02-96ER40989.

\end{document}